\documentstyle[12pt]{article}
\def\be{\begin{equation}}
\def\ee{\end{equation}}
\def\bea{\begin{eqnarray}}
\def\eea{\end{eqnarray}}

\begin{document}

\begin{center}
{\Large{\bf Two Moving-Angled 1-Branes with Electric Fields in a Partially Compact Spacetime}}

\vskip .5cm {\large Davoud Kamani and Niloufar Nowrouzi} \vskip
.1cm {\it Faculty of
Physics, Amirkabir University of Technology (Tehran Polytechnic)\\
P.O.Box: 15875-4413, Tehran, Iran}\\
{\it e-mails: n\underline{ }nowrouzi,
kamani@aut.ac.ir}\\
\end{center}

\begin{abstract}

In this article we consider two $m1$-branes at angle 
in the presence of the background electric fields, in a
partially compact spacetime. 
The branes have motions along a common direction that is
perpendicular to both of them. Using the boundary state
formalism, we calculate their interaction amplitude. Some special
cases of this interaction will be studied in detail.

\end{abstract}
%\vskip .5cm



\newpage

\section{Introduction}

In 1995 it is realized that open
strings with Dirichlet boundary conditions can end
on D-branes \cite{1}. Two D-branes which open string is stretched between
them can interact. It is known that one of the best methods
for finding some properties is calculation of the amplitude of
interaction. This amplitude was obtained by one loop open
string diagram. However, this is equivalent
to a tree-level diagram in the closed string exchange \cite{2}. A closed
string is generated from the vacuum,
propagates for a while and then annihilates again in the vacuum.
The state which describes the creation (annihilation) 
of closed string from (in) the
vacuum is called boundary state \cite{3}. So the boundary state
formalism is a strong tool for calculating the amplitude of interaction
of the branes, $e.g.$ see \cite{4,5,6,7,8,9,10,11,12,13,14,15} and
references therein.

In the case of D$p$-branes with nonzero background and internal
gauge fields, the boundary state formalism is an effective
method for calculating the amplitude of their interaction.
For a closed string emitted (absorbed) by a D$p$-brane
in the presence of the background field $B_{\mu\nu}$ and $U(1)$ gauge field
$A_\alpha$ (which lives on the brane), there are mixed boundary conditions.
This D$p$-brane is called $mp$-brane \cite{12,13,14,15}.

Previously we studied the interaction of two stationary $m1$-branes at angle
\cite{14}. In addition, we considered moving mixed branes \cite{15}. 
For both cases spacetime is compact. In this article we shall study both
cases simultaneously, i.e. a system of moving and angled
$m1$-branes in the partially compacted spacetime on a torus. At first the
boundary state associated with a moving $m1$-brane, which makes an angle
with the $X^1$-direction, will be obtained. It is parallel to the
$X^1X^2$-plane, and contains an electrical field along itself.
Then the interaction amplitude of the system $m1-m1'$ branes
will be obtained. The angle between the branes is $\phi$. The
branes move along the $X^3$-direction with the velocities $V_1$
and $V_2$. Various properties of the interaction
amplitude of this system will be analyzed. The large distance
behavior of the amplitude, which reveals the contribution of the
closed string massless states on the interaction, will be obtained.

This paper is organized as follows. In the section 2, we obtain
the boundary state corresponding to an oblique moving $m1$-brane.
In section 3, we obtain the amplitude of interaction via the
overlapping of two boundary states. In section 4, we suppose these
$m1$-branes are located at large distance. Thus, the contribution of
the massless states on the interaction will be studied. Section 5
is devoted to the conclusions.
%%%%%%%%%%%%%%%%%%%%%%%%%%%%%%%%%%%%%%%%%%%%%%%%%%%%%%%%%%%%%%
\section{Boundary state of an oblique moving $m1$-brane}

We suppose that an $m1$-brane with the electric field $E_2$ along
it, moves with the velocity $V_2$ along the direction $X^3$, while
makes an angle $\theta_2$ with the $X^1$-direction. It is parallel
to the $X^1X^2$-plane. In our notations the index 2 in $E_2$, $V_2$,
$\cdot \cdot \cdot$,
refers to the second $m1$-brane. Similarly, we consider $E_1$, $V_1$,
$\cdot \cdot \cdot$,
for the first $m1$-brane. Note that in this article the signature
of the metric is $\eta_{\mu\nu}=diag (-1,1, \cdot \cdot \cdot,1)$.

Previously we obtained the boundary state for
a moving $mp$-brane \cite{15}. In the corresponding boundary state 
equations we consider
$p=1$, and then rotate the $m1$-brane to make an angle $\theta_2$
with the $X^1$-direction. After this process the boundary state
equations, associated with the moving-angled $m1$-brane, take the
form
\begin{equation}
[\partial_{\tau}X^0-V_2\partial_{\tau}X^3-E_2\cos\theta_2\partial_{\sigma}
X^1-E_2\sin\theta_2\partial_{\sigma}X^2]_{\tau_0}|{B_x}^2,\tau_0\rangle=0,
\end{equation}
\begin{equation}
[\cos\theta_{2}\partial_{\tau}X^1+\sin\theta_2\partial_{\tau}X^2
-E_2\partial_{\sigma}X^0]_{\tau_0}|{B_x}^2,\tau_0\rangle=0,
\end{equation}
\begin{equation}
[X^3-V_2X^0-{y^3}_{(2)}]_{\tau_0}|{B_x}^2,\tau_0\rangle=0,
\end{equation}
\begin{equation}
[-(X^1-y^1_{(2)})\sin\theta_2+(X^2-y^2_{(2)})\cos\theta_2]_{\tau_0}
|{B_x}^2,\tau_0\rangle=0,
\end{equation}
\begin{equation}
(X^j-y^j_{(2)})_{\tau_0}|B^2_x,\tau_0\rangle=0,\quad\quad\quad
j\neq0,1,2,3.
\end{equation}

The mode expansion of $X^{\mu}(\sigma,\tau)$ is
\begin{equation}
X^{\mu}(\sigma,\tau)=x^{\mu}+2\alpha'p^{\mu}\tau+2L^{\mu}\sigma+
\frac{i}{2}\sqrt{2\alpha'}\sum_{m\neq0}\frac{1}{m}
(\alpha_m^{\mu}e^{-2im(\tau-\sigma)}
+\tilde{\alpha}_m^{\mu}e^{-2im(\tau+\sigma)}),
\end{equation}
where $L^\mu$ is zero for the non-compact directions. For a compact
direction there are $L^\mu=N^\mu R^\mu$ and $p^\mu=\frac{M^\mu}{R^\mu}$,
where $N^\mu$ and $M^\mu$ are winding number and momentum number
of the emitted (absorbed) closed string from the brane,
respectively. $R^\mu$ also is the radius of compactification of
the compact direction $X^\mu$.

After replacing the mode expansion of $X^\mu$ into the Eqs. (1)-(5) these
equations will be written in terms of the oscillators. 
The zero mode part of the boundary state equations become
\begin{equation}
[p^0-V_2p^3-\frac{1}{\alpha'}E_2(L^2\sin\theta_2+
L^1\cos\theta_2)]_{op}|{B_x}^2,\tau_0\rangle=0,
\end{equation}
\begin{equation}
[p^1\cos\theta_2+p^2\sin\theta_2-\frac{1}{\alpha'}E_2L^0]_{op}
|{B_x}^2,\tau_0\rangle=0,
\end{equation}
\begin{equation}
[-(x^1-y^1_{(2)}+2\alpha'\tau_0p^1)\sin\theta_2+
(x^2-y^2_{(2)}+2\alpha'\tau_0p^2)\cos\theta_2]_{op}|{B_x}^2,\tau_0\rangle=0,
\end{equation}
\begin{equation}
[L^2\cos\theta_2-L^1\sin\theta_2]_{op}|{B_x}^2,\tau_0\rangle=0,
\end{equation}
\begin{equation}
[x^3+2\alpha'\tau_0p^3-y^3_{(2)}-
V_2(x^0+2\alpha'\tau_0p^0)]_{op}|{B_x}^2,\tau_0\rangle=0,
\end{equation}
\begin{equation}
(L^3-V_2L^0)_{op}|{B_x}^2,\tau_0\rangle=0,
\end{equation}
\begin{equation}
[x^j+2\alpha'p^j\tau_0-y^j_{(2)}]_{op}|{B_x}^2,\tau_0\rangle=0,
\end{equation}
\begin{equation}
(L^j)_{op}|{B_x}^2,\tau_0\rangle=0.
\end{equation}
For the oscillating part, the equations of the boundary state are as in
the following
\begin{eqnarray}
&~&[(\alpha^0_m-V_2\alpha^3_m+E_2(\alpha^2_m\sin\theta_2+
\alpha^1_m\cos\theta_2))e^{-2im\tau_0}
\nonumber\\
&~&+(\tilde{\alpha}^0_{-m}-V_2\tilde{\alpha}^3_{-m}-
E_2(\tilde{\alpha}^2_{-m}\sin\theta_2+\tilde{\alpha}^
1_{-m}\cos\theta_2))e^{2im\tau_0}]|{B_x}^2,\tau_0\rangle=0,
\end{eqnarray}
\begin{eqnarray}
&~&[(E_2\alpha^0_m+\alpha^1_m\cos\theta_2+
\alpha^2_m\sin\theta_2)e^{-2im\tau_0}+
\nonumber\\
&~&(-E_2\tilde{\alpha}^0_{-m}+\tilde{\alpha}^1_{-m}\cos\theta_2+
\tilde{\alpha}^2_{-m}\sin\theta_2)e^{2im\tau_0}]|{B_x}^2,\tau_0\rangle=0,
\end{eqnarray}
\begin{equation}
[(-\alpha^1_m\sin\theta_2+\alpha^2_m\cos\theta_2)e^{-2im\tau_0}
+(\tilde{\alpha}^1_{-m}\sin\theta_2-
\tilde{\alpha}^2_{-m}\cos\theta_2)
e^{2im\tau_0}]|{B_x}^2,\tau_0\rangle=0,
\end{equation}
\begin{equation}
[(-V_2\alpha^0_m+\alpha^3_m)e^{-2im\tau_0}+
(V_2\tilde{\alpha}^0_{-m}-\tilde{\alpha}^3_{-m})
e^{2im\tau_0}]|{B_x}^2,\tau_0\rangle=0,
\end{equation}
\begin{equation}
[\alpha^j_me^{-2im\tau_0}-\tilde{\alpha}^j_{-m}
e^{2im\tau_0}]|{B_x}^2,\tau_0\rangle=0\quad,\quad
j\in\{4,\cdots,d-1\}.
\end{equation}
These equations can be collected in a single equation, i.e.,
\begin{equation}
({\alpha^\mu_m}e^{-2im\tau_0}+{{{S_{(2)}}^\mu}_\nu}\tilde{\alpha}^
\nu_{-m}e^{2im\tau_0})|{B_x}^2,\tau_0\rangle=0,
\end{equation}
where the matrix ${{S_{(2)}}^\mu}_\nu$ is defined by
\begin{equation}
{{S_{(2)}}^\mu}_\nu=\left(\begin{array}{l}
\vspace{0.5cm}
{\Omega_{(2)}}^p_{\;\;\;q}\quad\quad\quad\quad\quad 0 \\
\vspace{0.5cm}
0\quad\quad\quad -I_{(d-4)\times(d-4)}
\end{array}\right)\quad\quad,\quad\quad p,q\in\{0,1,2,3\}.
\end{equation}
The matrix ${\Omega_{(2)}}^p_{\;\;\;q}$ also has the definition
\begin{equation}
\scriptsize{\Omega_{(2)}}^p_{\;\;\;q} =\frac{1}{1-V^2_2-E^2_2} \left[
\begin{array}{llll}
\vspace{0.5cm}
1+V^2_2+E^2_2&-2E_2\cos\theta_2&-2E_2\sin\theta_2&-2V_2\\
\vspace{0.5cm}
-2E_2\cos\theta_2&(1-V^2_2)\cos2\theta_2+E^2_2&(1-V_2)
\sin2\theta_2&2V_2E_2\cos\theta_2\\
\vspace{0.5cm}
-2E_2\sin\theta_2&(1-V^2_2)\sin2\theta_2&-[(1-V^2_2)
\cos2\theta_2-E^2_2]&2V_2E_2\sin\theta_2\\
\vspace{0.5cm}
 2V_2&-2V_2E_2\cos\theta_2&-2V_2E_2\sin
\theta_2&-(1-E^2_2+V^2_2)
\end{array}\right].
\end{equation}
According to $({\Omega_{(2)}}^T)^p_{\;\;\;q}=\eta^{pp}\eta_{qq}{\Omega_{(2)}}^q_{\;\;\;p}$
the matrix $\Omega_{(2)}$ is orthogonal, and hence $S_{(2)}$ also is an orthogonal matrix.

By solving the Eqs. (7)-(14) and (20) the boundary state will be obtained
\begin{eqnarray}
&~&|{B_x}^2,\tau_0\rangle=\frac{T}{2}
\sqrt{1-V^2_2-E^2_2}\exp[i\alpha'\tau_0
(\gamma^2_2(p^3_{op}-V_2p^0_{op})^2
\nonumber\\
&~&+(-p^1_{op}\sin\theta_2+p^2_{op}\cos\theta_2)^2+
\sum_{j=4}^{d-1}(p^j_{op})^2)]
\nonumber\\
&~&\times\delta[-(x^1-y^1_{(2)})\sin\theta_2+
(x^2-y^2_{(2)})\cos\theta_2]\delta(x^3-y^3_{(2)}-V_2x^0)
\prod_{j=4}^{d-1}\delta(x^j-y^j_{(2)})
\nonumber\\
&~&\times\sum_{p^0}\sum_{p^1}\sum_{p^2}
|p^0\rangle|p^1\rangle|p^2\rangle
\prod_{j=4}^{d-1}|p^j_L=p^j_R=0\rangle
|p^3_L=p^3_R=\frac{1}{2}V_2p^0\rangle
\nonumber\\
&~&\times\exp[-\sum_{m=1}^\infty(\frac{1}{m}e^{4im\tau_0}
\alpha^\mu_{-m}S^{(2)}_{\mu\nu}\tilde{\alpha}^\nu_{-m})]|0\rangle,
\end{eqnarray}
where $\gamma_2=1/\sqrt{1-V_2^2}$, and
$T=\frac{\sqrt{\pi}}{2^{(d-10)/4}}(4\pi^2\alpha')^{(d-6)/4}$ is
tension of the $m1$-brane which lives in the $d$-dimensional
spacetime. The momentum components of the closed string, that are
appeared in (23), are given by
\begin{equation}
p^0=\frac{\gamma^2_2}{\alpha'}E_2(\ell^2\sin\theta_2+\ell^1\cos\theta_2),
\end{equation}
\begin{equation}
p^1=\frac{E_2}{\alpha'}\ell^0\cos\theta_2,
\end{equation}
\begin{equation}
p^2=\frac{E_2}{\alpha'}\ell^0\sin\theta_2,
\end{equation}
\begin{equation}
p^3=\frac{\gamma^2_2V_2}{\alpha'}E_2(\ell^2\sin\theta_2+\ell^1\cos\theta_2),
\end{equation}
where $p^\mu=p^\mu_L+p^\mu_R$ and
$\ell^\mu=\alpha'(p^\mu_L-p^\mu_R)=N^\mu R^\mu$. We should
consider (24)-(26) for summing over $p^0,p^1$ and $p^2$ in (23).
Therefore, these summations convert to the winding numbers $N^0,N^1$ and
$N^2$. The Eq. (24) implies that energy of the closed string is
quantized and depends on its winding numbers around the $X^1$ and
$X^2$ directions. However, the Eqs. (24)-(27) imply that 
the momentum numbers of the closed
string $M^0,M^1,M^2$ and $M^3$ are related to its winding numbers
$N^0,N^1$ and $N^2$. 

The Eqs. (10), (12) and (14) also lead to the relations
\begin{equation}
\ell^2\cos\theta_2=\ell^1\sin\theta_2,
\end{equation}
\begin{equation}
\ell^3=V_2\ell^0,
\end{equation}
\begin{equation}
\ell^j=0.
\end{equation}
We can write the Eq. (28) in the form
\begin{equation}
N^2R^2\cos\theta_2=N^1R^1\sin\theta_2.
\end{equation}
This equation tells us that only when
$\frac{R^1\sin\theta_2}{R^2\cos\theta_2}$ is rational, closed
string can wrap around $X^1$ and $X^2$ directions, otherwise
$N^1=N^2=0$ and closed string has no winding around $X^1$ and
$X^2$. In this case its energy also is zero. 
In the same way, by the Eq. (29), for having 
winding around $X^3$ and $X^0$, the quantity
$\frac{V_2R^0}{R^3}$ also should be rational.

The ghost part of the boundary state is independent of the
electric field $E_2$, the velocity $V_2$ and 
the angle $\theta_2$. It is given by
\begin{equation}
|B_{gh},\tau_0\rangle=\exp\bigg{[}\sum_{m=1}^\infty
e^{4im\tau_0}(c_{-m}{\tilde{b}}_{-m}-b_{-m}
\tilde{c}_{-m})\frac{c_0+\tilde{c}_0}{2}\bigg{]}
|q=1\rangle|\tilde{q}=1\rangle.
\end{equation}
%%%%%%%%%%%%%%%%%%%%%%%%%%%%%%%%%%%%%%%%%%%%%%%%%%%%%%%%%%%%%%%
\section{Interaction between two $m1$-branes}

Before calculation of the interaction amplitude, let us introduce
some notations for the positions of these two mixed branes. 
Similar to the $m1$-brane, the $m1'$-brane also is parallel 
to the $X^1X^2$-plane and makes angle
$\theta_1$ with the $X^1$-direction, and moves with the speed $V_1$
along the $X^3$-direction. The electric field on it also is $E_1$. 
The common direction of motions is $X^3$, and the other directions
perpendicular to the world-volume of both branes are $\{X^j| j \neq 0,1,2,3\}$. 
We use the set $\{X^{j_n}\}$ to denote the non-compact part of $\{X^j\}$, 
and $\{X^{j_c}\}$ is for the compact part of $\{X^j\}$.

Now we can calculate the overlap of the two boundary states to
obtain the interaction amplitude of the branes. The complete
boundary state for each brane is 
\be 
|B\rangle=|B_x\rangle|B_{gh}\rangle. 
\ee
These two mixed branes simply interact via exchange of
closed strings so the amplitude is given by 
\be 
{\cal A}=~^{^{(1)}}\langle B,\tau_0=0|D|B,\tau_0=0\rangle^{(2)}, 
\ee
where ``$D$'' is the closed string propagator. The calculation is
straightforward but tedious. Here we only write the final result
\begin{eqnarray}
&~&{\cal A}=\frac{T^2\alpha' L}{4(2\pi)^{d-4}|\sin\phi|
|V_1-V_2|}\sqrt{(1-V_1^2-E_1^2)(1-V_2^2-E_2^2)}
\nonumber\\
&~&\times\int_0^\infty dt
\bigg{\{}e^{(d-2)t/6}\bigg{(}\sqrt{\frac{\pi}{\alpha't}}\bigg{)}^{d_{j_n}}
\prod_{j_n}\exp \bigg{(}-\frac{(y^{j_n}_{(1)}-y^{j_n}_{(2)})
^2}{4\alpha't}\bigg{)}\prod_{j_c}\Theta_3 \bigg{(}\frac{y^{j_c}_
{(1)}-y^{j_c}_{(2)}}{2\pi
R_{j_c}}\bigg{|}\frac{i\alpha't}{\pi(R_{j_c})^2}\bigg{)}
\nonumber\\
&~&\times\sum_{N^0}\sum_{N^1}\sum_{N^2}\bigg{[}\exp[-\frac{t}{\alpha'}
(\ell^0\ell^0+(\ell^1\cos\theta_1+\ell^2\sin\theta_1)
(\ell^1\cos\theta_2+\ell^2\sin\theta_2)
\nonumber\\
&~&+F^{(+)}F^{(-)})+\frac{i}{\alpha'}
(\Phi(12)y^3_{(2)}-\Phi(21)y^3_{(1)})]\bigg{]} \Theta_3(\nu|\tau)
\nonumber\\
&~&\times \prod_{n=1}^\infty[\det(1-\Omega_1
\Omega_2^Te^{-4nt})]^{-1}(1-e^{-4nt})^{6-d}\bigg{\}},
\end{eqnarray}
where $L=2\pi R_0$, and $\Phi(12)$ and $F^{(\pm)}$ are defined by
\begin{eqnarray}
&~&\Phi(12)=\frac{1}{V_2-V_1}[\gamma^2_1E_1(V^2_1+1)(\ell^2\sin
\theta_1+\ell^1\cos\theta_1)
\nonumber\\
&~&-\gamma^2_2E_2(1+V_1V_2)(\ell^2\sin\theta_2+\ell^1\cos\theta_2)],
\end{eqnarray}
\begin{eqnarray}
&~&F^{(\pm)}=\frac{1}{|V_1-V_2|}[\gamma^2_2(1\pm
V_1)(1+V_2^2)E_2 (\ell^2\sin\theta_2+\ell^1\cos\theta_2)
\nonumber\\
&~&-\gamma^2_1(1\pm
V_2)(1+V^2_1)E_1(\ell^2\sin\theta_1+\ell^1\cos\theta_1)].
\end{eqnarray}
We can obtain $\Phi(21)$ by exchanging $1\longleftrightarrow 2$ in
(36). In addition, $\phi=\theta_2-\theta_1$ and $\nu$ and $\tau$
also have the definitions
\begin{eqnarray}
&~&\nu=\frac{R_0}{2\pi\alpha'\sin\phi}[(E_2-E_1\cos\phi)
\bar{y}^2_{(1)}+(E_1-E_2\cos\phi)\bar{y}^2_{(2)}],
\nonumber\\
&~&\tau=\frac{itR_0^2}{\pi\alpha'}\bigg{(}\frac
{E_1^2+E_2^2-2E_1E_2\cos\phi}{\sin^2\phi}-1\bigg{)}.
\end{eqnarray}
The set $\{\bar{y}_{(2)}^2,y_{(2)}^3,\cdots,y_{(2)}^{(d-1)}\}$ shows
the position of the $m1$-brane, with
$\bar{y}^2_{(2)}=-y^1_{(2)}\sin\theta_2+y^2_{(2)}\cos\theta_2$
and $y^1_{(2)}\cos\theta_2+y^2_{(2)}\sin\theta_2=0$, similarly
for the $m1'$-brane. We observe that the interaction
amplitude not only depends on the relative angle $\phi$ between
the branes, but also depends on the configuration angles of the
branes, i.e. $\theta_1$ and $\theta_2$.

Because of the electric fields, this amplitude is not symmetric under
the change $\phi\rightarrow \pi-\phi$. Therefore, for the angled
mixed branes, $\phi$ and $\pi-\phi$ indicate two different
configurations. From (38), we see that the electric fields and
compactification of the time direction cause $\bar{y}^2_{(2)}$ and
$\bar{y}^2_{(1)}$ to appear in the interaction. Finally, the
amplitude (35) is symmetric with respect to the $m1$ and $m1'$-
branes, i.e.,
\begin{equation}
{\cal A}(V_1,V_2;E_1,E_2;\theta_1,\theta_2;y_1,y_2)=
{\cal A}^*(V_2,V_1;E_2,E_1;\theta_2,\theta_1;y_2,y_1).
\end{equation}
For complex conjugation see (34).

For non-compact spacetime, remove all factors
$\Theta_3$ from (35). In addition, use $\ell^0=\ell^1=\ell^2=0$,
and change $j_n\rightarrow j$, and hence $d_{j_n}\rightarrow d-4$.
So the interaction amplitude in the non-compact spacetime is
as in the following
\begin{eqnarray}
&~&{\cal A}_{\rm non-compact}=\frac{T^2\alpha' L}{4(2\pi)^{d-4}
|\sin\phi||V_1-V_2|}
\sqrt{(1-V^2_1-E^2_1)(1-V^2_2-E^2_2)}
\nonumber\\
&~&\times\int_0^\infty
dt\bigg{\{}(e^{(d-2)t/6}\bigg{(}\sqrt{\frac{\pi}
{\alpha't}}\bigg{)}^{d-4}\exp \bigg{(}-\sum_{j=4}^{d-1}\frac
{(y^j_{(1)}-y^j_{(2)})^2}{4\alpha't}\bigg{)}
\nonumber\\
&~&\times\prod_{n=1}^\infty[(\det(1-\Omega_1
\Omega^T_2e^{-4nt}))^{-1}(1-e^{-4nt})^{6-d}])\bigg{\}}.
\end{eqnarray}
This interaction depends on the minimal distance between the branes,
that is $\sum_{j=4}^{d-1}(y^j_{(1)}-y^j_{(2)})^2$.
%%%%%%%%%%%%%%%%%%%%%%%%%%%%%%%%%%%%%%%%%%%%%%%%%%%%%%%%%%%%%%%%%
\section{Large distance branes}

Now we extract the contribution of the massless states in the
interaction. As the metric $G_{\mu\nu}$, anti-symmetric tensor
$B_{\mu\nu}$ and dilaton $\Phi$ have zero winding and zero
momentum numbers, only the term with $N^0=N^1=N^2=0$ corresponds
to these massless states. By using the identity $det M=e^{Tr(\ln
M)}$ for a matrix M, we obtain the following limit for $d=26$
\begin{eqnarray}
&~&\lim_{q\rightarrow0}\frac{1}{q}\prod_{n=1}^\infty
\bigg{(}[\det(1-\Omega_1\Omega_2^T
q^n)]^{-1}(1-q^n)^{-20}\bigg{)}
\nonumber\\
&~&=\lim_{q\rightarrow0}\frac{1}{q}+{\rm Tr}(\Omega_1\Omega_2^T)+20,
\end{eqnarray}
where $q=e^{-4t}$. Put away the tachyon divergence, the
contribution of the massless states is given by
\begin{eqnarray}
&~&{\cal A}^{(0)}=\frac{T^2\alpha'L} {4(2\pi)^{22}|\sin\phi||V_1-V_2|}
\sqrt{(1-V^2_1-E^2_1)(1-V_2^2-E_2^2)}
[{\rm Tr}(\Omega_1\Omega_2^T)+20]G,
\nonumber\\
&~&G\equiv\int_o^\infty
dt\bigg{\{}\bigg{(}\sqrt{\frac{\pi}{\alpha't}}\bigg{)}^{d_{j_n}}
\prod_{j_n}\exp \bigg{(}-\frac{(y^{j_n}_{(1)}-y^{j_n}_{(2)})^2}{4\alpha't}
\bigg{)} \prod_{j_c}\Theta_3 \bigg{(}\frac{y^{j_c}_{(1)}-y^{j_c}_{(2)}} {2\pi
R_{j_c}}\bigg{|}\frac{i\alpha't}
{\pi(R_{j_c})^2}\bigg{)}\Theta_3(\nu|\tau)\bigg{\}}.
\end{eqnarray}
For the non-compact spacetime this amplitude reduces to
\begin{equation}
{\cal A}^{(0)}_{\rm non-compact}=\frac{T^2 \alpha' L}{4(2\pi)^{22}
|\sin\phi||V_1-V_2|} \sqrt{(1-V^2_1-E_1^2)(1-V_2^2-E_2^2)}
[{\rm Tr}(\Omega_1\Omega_2^T)+20]G_{22}({\bar Y}^2),
\end{equation}
where $\bar{Y}^2=\sum_{j=4}^{25}(y^j_1-y^j_2)^2$ is the impact
parameter, and $G_{22}$ is the Green's function of 
the 22-dimensional space.
%%%%%%%%%%%%%%%%%%%%%%%%%%%%%%%%%%%%%%%%%%%%%%%%%%%%%%%%
\section{Conclusions}

We obtained the boundary state, associated with an oblique moving $m1$-
brane, parallel to the $X^1X^2$-plane. This state reveals that
how electric field, velocity of the brane, obliqueness of the brane,
compact part and non-compact part of
the spacetime affect the brane. For a closed string
emitted (absorbed) by such brane, some of the momentum numbers have
relations with the winding numbers.

We determined the interaction amplitude of two
moving-angled $m1$-branes, which live in the partially compact
spacetime. This amplitude depends on the electric fields, 
velocities of the branes, obliqueness of the branes,
compact part and non-compact part of the spacetime. In addition,
this interaction contains the relative angle $\phi$,
and configuration angle of each brane, i.e. $\theta_1$ and
$\theta_2$. The electric fields along the branes imply that the cases
$\phi$ and $\pi-\phi$ are two different systems. 

We extracted contribution of the massless states (i.e. graviton,
dilaton and Kalb-Ramond fields) on
the interaction. For the non-compact spacetime, this contribution
is proportional to the Green's function of the 22-dimensional space.
%%%%%%%%%%%%%%%%%%%%%%%%%%%%%%%%%%%%%%%%%%%%%%%%%%%%%%%%%%%%%%%%%%%

\end{document}